# Electrocardiogram Heartbeat Classification Using Convolutional Neural Networks for the Detection of Cardiac Arrhythmia

Mohammad Mahmudur Rahman Khan[1], Md. Abu Bakr Siddique[2], Shadman Sakib[3], Anas Aziz[4], Abyaz Kader Tanzeem[5], Ziad Hossain[6]
[1]Department of ECE, Vanderbilt University, Nashville, Tennessee-37240, USA
[2]Department of EEE, International University of Business Agriculture and Technology, Dhaka-1230, Bangladesh
[3]Department of EECS, University of Hyogo, 2167 Shosha, 671-2280, JAPAN
[4]Department of AE, Military Institute of Science and Technology, Dhaka-1216, Bangladesh
[5]Department of EEE, BRAC University, Dhaka-1212, Bangladesh
[6]Department of CSE, North South University, Dhaka-1229, Bangladesh
mohammad.mahmudur.rahman.khan@vanderbilt.edu[1], absiddique@iubat.edu[2], sakibshadman15@gmail.com[3],
anas.aziz@ae.mist.ac.bd[4], tanzeemabyaz@gmail.com[5], ziadhossain98@gmail.com[6]

*Abstract*—The classification of the electrocardiogram (ECG) signal has a vital impact on the identification of heart-related diseases. This can ensure the premature finding of heart disease and the proper selection of the patient's customized treatment. However, the detection of arrhythmia is a challenging task to perform manually. This justifies the necessity of a technique for automatic detection of abnormal heart signals. Therefore, our work is based on the classification of five classes of ECG arrhythmic signals from Physionet's MIT-BIH Arrhythmia Dataset. Artificial Neural Networks (ANN) have demonstrated significant success in ECG signal classification. Our proposed model is a Convolutional Neural Network (CNN) customized for the categorization of the ECG signals. Our result testifies that the planned CNN model can successfully categorize arrhythmia with an overall accuracy of 95.2%. The average precision and recall of the proposed model are 95.2% and 95.4% respectively. This model can effectively be used to detect irregularities of heart rhythm at an early stage.

## I. INTRODUCTION

Cardiac conduction system of the human heart is accountable for the generation and control of heartbeats. It does so by repolarization followed by depolarization of the atrial and ventricular cardiomyocytes which can also be referred to as the electrical activity within the heart. Problems within the cardiac conduction system can introduce aberrations in the electrical impulses which overturn the normal behavior of speed as well as the rhythm of the heartbeat. This medical condition is broadly termed as Cardiac Dysrhythmia or Arrhythmia. Irregularity in heartbeat rhythm (Fibrillation), heart rate over 100 bpm (Tachycardia) and below 60 bpm (Bradycardia) are known as the types of arrhythmia. Medical studies show that the long-term effects of arrhythmia can range from being harmless to lethal depending on its type [1]. 15-20% of the deaths occurring worldwide has cardiac dysrhythmia and sudden cardiac death as the underlying cause [2]. However, an early-stage diagnosis could result in appropriate treatment and an eventual cure.

An extremely suitable medical tool for visualizing the heart's electrical activity is the Electrocardiogram (ECG). Apart from being non-invasive, it is also highly efficient, fast and easy-to-use. A heartbeat is represented as three major waves on the ECG display which are named as P-wave, QRS-complex and T-wave. Therefore, the presence of arrhythmia in a patient is detected with the proper judgment of the obtained ECG waveform by an expert cardiologist. For detecting some types of arrhythmia, each heartbeat's data over a long period of time is first recorded using Holter and loop recorder [3] and then needs to undergo careful visual analysis as the abnormality does not appear frequently. Manual completion of this task becomes very tiresome and lingering for the expert. Thus, the use of automatic methods has become crucial.

Previously, numerous automatic ECG data classification approaches using techniques such as hidden Markov models [4], wavelet transforms [5], support vector machine [6], Artificial Neural Network [7] etc. were developed. Feature extraction as well as signal pre-processing was a crucial requirement for these techniques to be applied. Extracting the features required the involvement of a medical expert and was done using hand-crafted methods. Therefore, these techniques became time-consuming, expensive and susceptible to the loss of data in the feature extraction phase. Additionally, these techniques faced a lot of significant challenges due to the morphological features of the signal having the nature of being highly individual and variable i.e. same symptoms of arrhythmia may display different morphologies of the signal in varying circumstances. Hence, a good classification performance could not be achieved when exposed to new ECG data.

Ye *et al.* [8] combined a general multi-class classifier (incremental SVM) with a specific classifier (two-class SVM) and obtained an improved performance for ECG signal classification of heartbeats with an accuracy of 86%. Classification of ECG waveform using RS along with QNN was done by Tang et al. [9] where an overall accuracy of 91.7 % was obtained. Most recently, however, CNN has gained immense fame in the field of audio [10] and image classification [11] due to its automatic feature learning, reduced computational complexity and hence fast classification capability. Additionally, biomedical applications [12], [13] are also being addressed by CNN currently. A patient-specific arrhythmia detection system from real-time ECG signal was developed using 1D deep CNN [14] which had the drawback of facing a lot of difficulties to be able to cope with real-world scenario since their training data set was composed of patient-specific ECG signals. However, Mattila et al. [15] developed an inter-patient ECG classifier using 1D CNN for the efficient finding of arrhythmia from 3-classes of heartbeats which was more realistic since their training dataset did not contain any test data. Besides, AlexNet coupled with a back-propagation neural network was implemented to detect dysrhythmia from the ECG waveform of three different heart conditions by Ali et al. [16] and an accuracy of 92% was achieved. Also, an accuracy of 92.70% was obtained by Zubair et al. [17] where 1D CNN was used to categorize 5 classes of heartbeats from ECG signals. Kachuee et al. [18] used a deep CNN approach for detection of arrhythmia and transfer this function to the detection of Myocardial infraction analyzing the ECG heartbeat signals and showed an accuracy of 93.40%. Additionally, Acharya et al. [19] used deep CNN to successfully identify arrhythmic heartbeats from the ECG data of 5 different classes of heartbeats obtaining an accuracy of 94.03%. Furthermore, Implementation of 2D CNN to detect dysrhythmia by classifying ECG signals was done by Zhai et al. [1] where both the individual characteristics of the beat along with the beat-to-beat temporal relationship was captured which made the performance far superior than the 1-D approaches. Another 2-D deep CNN classifier was developed [20] for the efficient detection of arrhythmia from ECG signals which was further optimized by a few deep learning techniques.

A 1-D CNN is implemented in this study for automated heartbeat classification of five different forms of cardiac dysrhythmia. Conventional ECG signal processing methods are instigated to eliminate noise from the data, to detect peaks, and to segment the heartbeat signal. Augmentation is applied to training data resulting in higher classification accuracy. The method is evaluated based on performance evaluation matrices where it produces the best outcomes compared to the current literature.

The next sections of the paper are ordered according to this. Section II describes the materials and the detailed method used for the classification of the ECG heartbeat comprising of a few subsections: the overview of the ECG database, pre-processing of the ECG signal, data augmentation, and the analysis of the proposed 1-D CNN ECG classifier. Section III presents the experimental results, evaluation, and comparison, and subsequently, section IV provides the conclusion and future works.

## II. MATERIALS AND METHODOLOGY

To classify the ECG heartbeat, a 1-D CNN has been utilized. Before that, several preprocessing has been applied to the ECG signal. The detailed steps are discussed in the subsequent section. Our overall proposed framework is presented in Figure 1.

### A. Data Acquisition

The MIT-BIH Arrhythmia [21] is a freely available dataset which has been extensively used to assess the performance of ECG based heartbeat categorization algorithms [14], [22]. This benchmark dataset comprises of 48 records of two channels ECG signals for 30 minutes duration collected from 47 individuals. In this paper, a total of 109446 beats at 125 Hz sampling frequency from 44 records are evaluated as train and test patterns for the performance analysis of 1D convolutional neural network model. 4 paced beats are kept out in the evaluation task since these beats do not preserve adequate signal features for sound processing. According to AAMI recommendations, each ECG beat can be categorized into 5 heartbeat types: N – Normal Beat, SVP – Supra-Ventricular Premature Beat, PVC – Premature Ventricular Contraction Beat, FVN – Fusion of Ventricular and Normal Beat, FPN – Fusion of Paced and Normal Beat. Figure 2 displays the five different arrhythmia heartbeat signal obtained from MIT-BIH Arrhythmia dataset.

### B. ECG Signal Preprocessing

There are different forms of preprocessing needed to enhance the efficiency of the ECG signal and to generate the ECG beats from a specified ECG waveform. A few key ways to pre-process the ECG database is adopted: (i) de-noising; (ii) peak detection of QRS; and (iii) segmentation of the heartbeat signals.

ECG signal includes different forms of destructive noise. Therefore, the first step then is to de-noise the ECG signal by removing the noises out of the signal. At first, for the reduction of the dc noise present in the ECG signals, mean filtering is implemented. Then the unnecessary dc module is eradicated by deducting the average of the ECG data from each test sample, and also the amplitude of the signal threshold is taken down to level zero. Nearly all ECG recordings generate high and low-frequency noise affected by innumerable reasons such as muscle activity, cardiac activities, improper interaction with electrodes, and the influence of many other environmental factors. However, due to these artifacts, the relevant data cannot be quickly derived from the raw signal, and therefore must be analyzed first to model and de-noise the ECG signal. In addition to lowering the artifacts, the signal is often normalized. In this research, the ECG signal is rescaled to the range [-1, 1]. The following equation used for rescaling the signal.

$$Y_n = Y_n / \max(Y) \qquad (1)$$

where $Y_n$ is the value of feature N and X is the array of the samples.

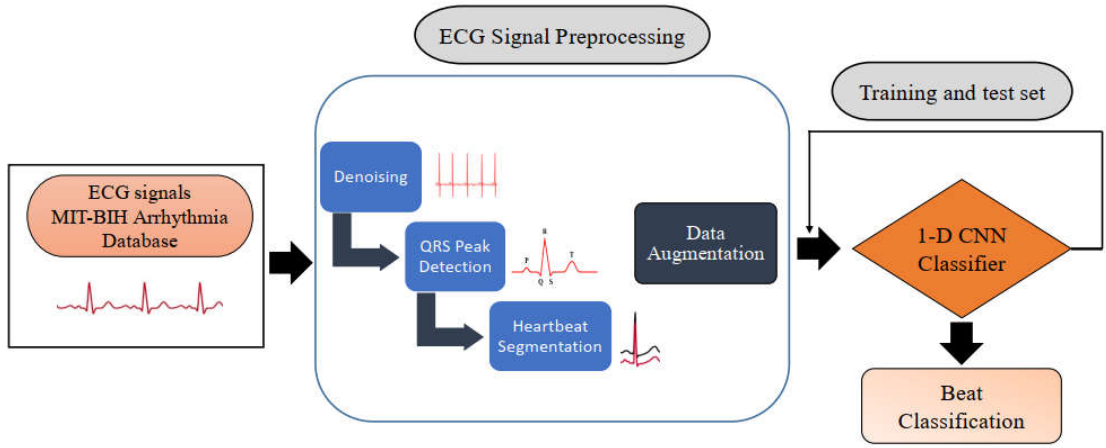

Fig. 1. Schematic design of the projected methodology for ECG heartbeat classification

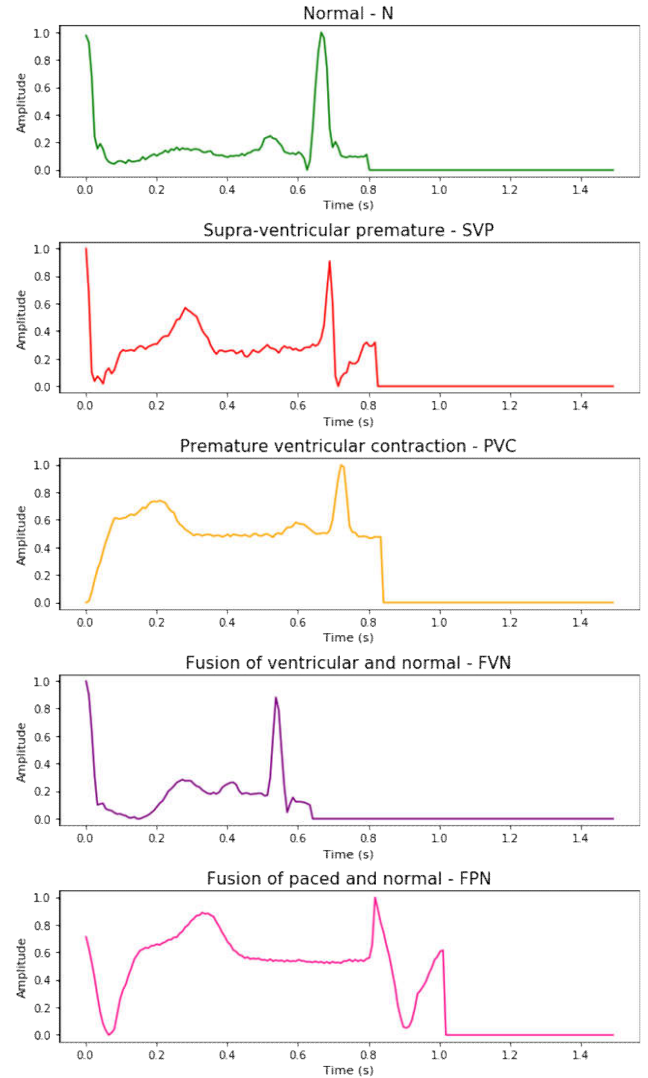

Once the signals have been processed, peak detection is done which is effective for segmenting the ECG signal into single beats. QRS is the most remarkable waveform within the ECG and forms the framework for almost all automatic ECG diagnostic algorithms. As it illustrates the electrical impulses inside the heart mostly during ventricular contraction, valuable data on the actual condition of the heart is given by the frequency of its existence along with its shape. Almost every R peak detection is equivalent to a single heartbeat detection. To perform the QRS detection a well-known Pan-Tompkins algorithm [23] is applied. The technique encompasses a series of mechanisms implementing derivative, squaring, integration, edge detection, and search techniques for identifying R-peaks of the ECG signal. Finally, after QRS detection, locating the P, R, and T peaks, the segmentation of individual heartbeat is performed.

### C. Data Augmentation and Splitting

To train the model properly, all the data should be augmented to the same level. For this, the smallest class samples of a heartbeat have been selected. Due to the imbalance of the dataset, which could result in misclassification, the data is down-sampled. All the samples are cropped, down-sampled, and padded with zero to the fixed dimension of 188 samples to illustrate the raw data of individual beats. Later, these samples are given into the input/starting layer of CNN.

After data augmentation, the number of samples in each heartbeat classes are equal. 800 samples are taken from each of the five heartbeat classes by splitting randomly which makes the test set of 4000 samples in total. The training set consists of 109150 samples with 21830 samples in individual heartbeat class. All the training heartbeats are 188 samples in length. Figure 3 shows the sample plotting of training heartbeat signals. However, as neural networks will be used for our classification model, one-hot encoding to turn our output classes is used into a numerical representation.

Fig. 2. Downsampled ECG arrhythmia heartbeat signal of five classes

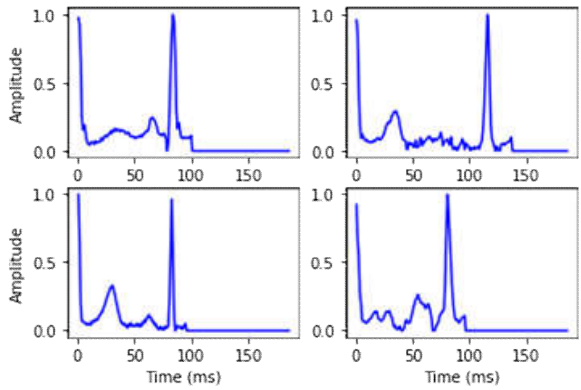

Fig. 3. Sample plot of the extracted training heartbeats of different class

*D. Proposed 1-D CNN ECG Heartbeat Classifier*

CNN is recognized as the feature learner that typically consists of two parts, and has a tremendous capacity to automatically draw out essential feature from input data. The first step is feature extractor, which involves a convolutional layer and pooling layers, and automatically learn the characteristics from raw data. Then fully connected layer executes the classification from the first part relying on the learned attributes. The input layer is composed of the individual values that denote the smallest unit of input whereas the output layer comprises as many outputs as categories exist in the particular classification problem. The convolutional layer performs an activity of convolution to limited localized regions through transforming a layer to the preceding layer. This is used specifically for extracting the feature from the raw data. Pooling layers are employed after convolutional layers, that minimizes the amount of parameters associated and minimize computational complexity.

In this research, a 1-D CNN is adopted as the ECG heartbeat signal classifier. CNN layers are utilized to automatically draw out an attribute from the ECG signal. Our proposed CNN structure comprised of 4 convolutional layers, three pooling layers afterwards a single fully connected layer or dense layer, and a softmax. Figure 4 illustrates the CNN framework for automated heartbeat classification.

The first layer is a 1-D convolutional layer consists of 32 filters with a kernel size of 5×1, a stride of 1 with ReLU activation function which simply returns the value provided if positive, else it returns 0. The layer following the convolutional layer is the max-pooling layer with a pool size of 5×1 and a stride of 1. It cuts down the number of parameters by half by only choosing the neurons with maximum activation value within a 5×1 region. The padding is kept the same which means that both the output and input feature maps have the matching spatial dimensions. A dropout of 0.25 is added followed by the max-pooling layer. It randomly sets 25% of the neurons to 0 (i.e. disables them). Doing so minimizes the problem of overfitting of the network. The second 1-D convolutional layer comprised of 64 filters with a kernel size of 5×1, a stride of 1 followed by a ReLU. Subsequently, the 3$^{rd}$ and 4$^{th}$ convolutional layers consist of 128 and 256 filters with a kernel size of 5×1, stride 1 followed by a ReLU. These convolutional layers are also ensuing by max-pooling layers with a pool size of 5×1. These layers are then followed by a flattening layer that converts the multi-dimensional feature vector into a 1-D feature vector preparing the output for a fully connected layer. The output of the flattening layer is then given into a fully connected or dense layer of 512 units followed by a ReLU. A fully connected layer binds each neuron from the layer from the preceding layer to the subsequent layer. It is then followed by a dropout layer to reduce overfitting. Finally, the softmax activation function is applied for the prediction of the class to which the input data belongs. The output size of this layer is 5 since having 5 classes to classify the ECG heartbeat signal.

III. RESULTS ANALYSIS AND DISCUSSION

For the automatic categorization of ECG heartbeat, MIT-BIH Arrhythmia dataset is utilized to assess the performance of the model using 1-D CNN. According to AAMI recommendations, each ECG beat can be categorized into the 5 heartbeat types: N – Normal Beat, SVP – Supra-Ventricular Premature Beat, PVC – Premature Ventricular Contraction Beat, FVN – Fusion of Ventricular and Normal Beat, FPN – Fusion of Paced and Normal Beat. The training set consists of 109150 beat samples while the testing set consists of 4000 beat samples in total each class with 800 samples. For training the network, Adam selecting the starting learning rate of 0.001, the decay rate of 0.75 for 60 epochs, and sparse categorical cross-entropy as a loss function is used. Our 1-D CNN model is deployed in Keras and Tensorflow GPU backend. The computer configuration was Intel Xeon E3 with 16GB RAM and NVIDIA Geforce GTX1080Ti GPU.

*A. ECG Heartbeat Categorization Performance Evaluation*

The CNN model, suggested in this paper, was implemented on the Physionet's MIT-BIH Arrhythmia dataset which consists of 5 types of ECG signals: Normal, Supra-ventricular premature, Premature ventricular contraction, Fusion of ventricular and normal and Fusion of paced and normal. These signals varied in their sample sizes to a great extend which made the dataset imbalanced. The sample sizes for training and testing are provided in Table 1.

TABLE I. TRAINING AND TESTING SAMPLE SIZES OF ECG BEATS

| Heartbeat Type | Training sample size | Testing sample size |
|---|---|---|
| Normal | 21830 | 800 |
| Supra-ventricular premature | 21830 | 800 |
| Premature ventricular contraction | 21830 | 800 |
| Fusion of ventricular and normal | 21830 | 800 |
| Fusion of paced and normal | 21830 | 800 |
| Total | 109150 | 4000 |

As the dataset is imbalanced, rather than solely relying on the classification accuracy as a model performance evaluating metric, precision, recall and F1-score evaluation are considered to justify the preeminence of the projected model. If a two-class classification problem is considered, the performance of a classifier model can be observed as a confusion matrix [24], illustrated in Table II.

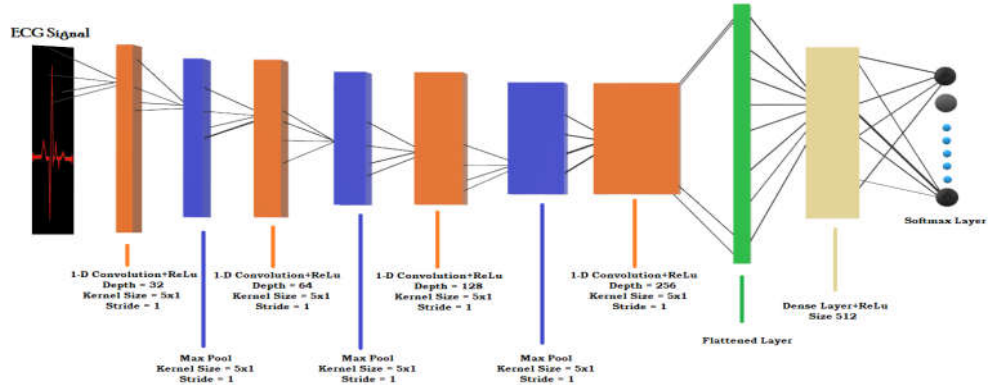

Fig. 4. Proposed 1-D CNN architecture for heartbeat classification

TABLE II. CONFUSION MATRIX

|  | Predicted Positive | Predicted Negative |
|---|---|---|
| Actual Positive | True Positives (TP) | False Negatives (FN) |
| Actual Negative | False Positives (FP) | True Negatives (TN) |

From the confusion matrix provided in Table II, the overall accuracy can be computed by Eqn. 2.

$$Accuracy(\%) = (TP + TN) / (TP + FP + TN + FN) \quad (2)$$

However, as mentioned earlier, the precision, recall and the F1-score values are also evaluated for analyzing the performance of the CNN model. The equations of these metrics are provided in Equation 2, 3 and 4 respectively. The precision value represents classifier model's exactness. On the other hand, the recall value represents the model's completeness.

$$Precision = TP / (TP + FP) \quad (3)$$

$$Recall = TP / (TP + FN) \quad (4)$$

$$F1-score = \frac{2 * Precision \times Recall}{Precision + Recall} \quad (5)$$

The corresponding values of the performance evaluating metrics are provided in Table III. The suggested CNN model performs well for identifying the premature ventricular contraction and the Fusion of paced and normal if evaluated based on the precision value. On the other hand, the model is good for detecting the Fusion of ventricular and normal if the recall value of the model is taken under consideration.

However, the model showed higher F1- score in case of the Fusion of paced and normal signals.

TABLE III. CLASSIFICATION PERFORMANCE OF THE PROJECTED CNN MODEL

| ECG Beat Type | Precision | Recall | F1-score |
|---|---|---|---|
| Normal | 0.97 | 0.90 | 0.93 |
| Supra-ventricular premature | 0.94 | 0.97 | 0.95 |
| Premature ventricular contraction | 0.99 | 0.94 | 0.96 |
| Fusion of ventricular and normal | 0.87 | 0.99 | 0.93 |
| Fusion of paced and normal | 0.99 | 0.97 | 0.98 |
| Weighted Average | 0.952 (95.2%) | 0.954 (95.4%) | 0.950 (95.0%) |

The testing performance of the model is demonstrated in the confusion matrix provided in Figure 5.

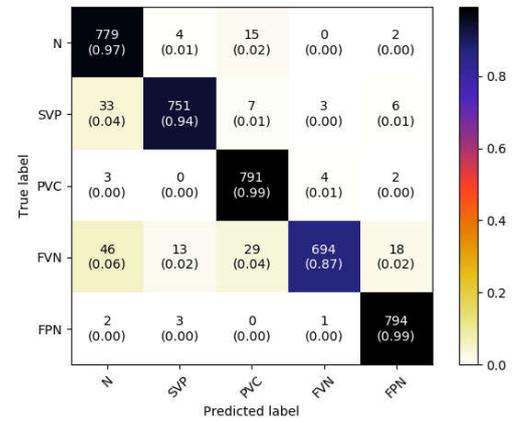

Fig. 5. Confusion matrix of the suggested 1-D CNN model on test data

Evaluating the confusion matrix, it is sensible to state that the CNN model performs very well classifying the ECG signals, specially, detecting the Fusion of paced and normal ECG signal. The overall accuracy of the model is 95.2% which justifies the effectiveness of the classification model.

### B. Comparison with the existing methods

To assess our classification, our proposed 1-D CNN ECG classifier is compared to the state-of-the-art methods. Table IV presents the performance comparison of our suggested model with previously developed algorithms. Our suggested model outperforms the existing approaches by demonstrating a classification accuracy of 95.2%.

TABLE IV. RESULT COMPARISON WITH THE EXISTING METHODS

| Methods | Overall Accuracy (%) |
|---|---|
| Ye et al. [8] | 86.0 |
| X. Tang et al. [9] | 91.7 |
| Ali et al. [16] | 92.0 |
| Zubair et al. [17] | 92.7 |
| Kachuee et al. [18] | 93.4 |
| Acharya et al. [19] | 94.0 |
| **Proposed method** | **95.2** |

## IV. CONCLUSION

In this study, a deep learning 1-D CNN is proposed for the automatic ECG heartbeat categorization to categorize five different types of cardiac arrhythmia. For better performance, the ECG signals were processed using several preprocessing steps (denoising, peak detection, heartbeat segmentation). The proposed ECG heartbeat classification systems performance was validated from Physionet's MIT-BIH Arrhythmia Dataset. Experimental results demonstrate that our suggested model achieved an overall classification accuracy of 95.2% with an average precision and recall of 95.2% and 95.4%. Thus our proposed deep learning framework significantly outperforms the previous state-of-art methods. Furthermore, the suggested ECG arrhythmia classifier can be applied in several biomedical applications such as sleep staging, a medical robot that monitors the ECG signal and assists the medical experts to detect cardiac arrhythmia more accurately. As part of our future work, our framework will be extended by implementing 2-D CNN with ECG greyscale image input which will be transformed from the MIT-BIH Arrhythmia Dataset ECG recording.